\documentclass[aps,reprint,twocolumn]{revtex4-2}
\usepackage[utf8]{inputenc}
\usepackage{amsmath,amssymb}
\usepackage{graphicx}
\usepackage{textgreek}

\begin{document}
	
\title{Laser micromachining of arbitrarily complex and overhang-free SiN nanomechanical resonators}
	
\author{Yahya Saleh\textsuperscript{1}}
\altaffiliation{These authors contributed equally to this work.}

\author{Zachary Louis-Seize\textsuperscript{1}}
\altaffiliation{These authors contributed equally to this work.}

\author{Timothy Hodges\textsuperscript{1}}

\author{David Girard\textsuperscript{2}}

\author{Mohammed Shakir\textsuperscript{1}}

\author{Mathis Turgeon-Roy\textsuperscript{1}}

\author{Francis Doyon-D'Amour\textsuperscript{1}}

\author{Chang Zhang\textsuperscript{1}}

\author{Arnaud Weck\textsuperscript{1,2}}

\author{Raphael St-Gelais\textsuperscript{1,2}}
\email{raphael.stgelais@uottawa.ca}

\affiliation{\textsuperscript{1}Department of Mechanical Engineering, University of Ottawa}
\affiliation{\textsuperscript{2}Department of Physics, University of Ottawa}

\date{\today}
	
\begin{abstract}
Research on silicon nitride (SiN) nanomechanical resonators produces an exceptionally rich variety of resonator geometries, for which there is currently no available rapid prototyping solution. Experimental advances in nanobeam, trampoline, phononic bandgap, and soft-clamping structures all rely on conventional nanofabrication involving e-beam or photolithography, followed by various etching steps. These techniques are typically time-consuming, relatively inflexible, and often result in spurious residual SiN overhang that can degrade mechanical quality factors. In contrast, recent work has shown that simple resonant structures, such as nanobeams, can be prototyped by direct laser ablation of free-standing SiN membranes using a spatially distributed sequence of microholes that limits stress concentration. However, these early demonstrations were restricted to basic shapes, created by manually combining ablation routines for circles and straight lines. Here, we demonstrate the fabrication of arbitrarily complex geometries using an open-source software toolset---released with this publication---that automatically generates laser-ablated hole sequences directly from standard semiconductor layout files (i.e., GDSII). The software includes a layout alignment tool that compensates for the membrane orientation and dimensional variations, limiting material overhang to $\sim2$ \textmu m. Using this toolset, we fabricate several resonator geometries, each in under 1 hour, two of which are exhaustively characterized as candidate structures for high-performance radiation sensing. The measured quality factors of these structures closely match finite element simulations and reach values up to $3.7\times10^6$. From these measurements, we extract material quality factors above 3700, which is on par with low-stress SiN unablated plain membranes and with comparable structures produced using conventional fabrication methods.
\end{abstract}

\maketitle
	
\section{Intro}
Silicon nitride (SiN) nanomechanical resonators have enabled a diverse range of devices for sensing, signal processing, and fundamental physics research. Trampoline resonators have been extensively studied for their high quality factors \cite{reinhardt_ultralow-noise_2016,norte_mechanical_2016} and thermal isolation \cite{piller_thermal_2023}, while nanobeams serve as the foundation for many optomechanical and sensing applications \cite{aspelmeyer_cavity_2014,verbridge_high_2006}. More complex architectures, such as phononic bandgap \cite{tsaturyan_demonstration_2014,clark_optically_2024}, hierarchical tensile \cite{bereyhi_hierarchical_2022}, and spiderweb-inspired structures with soft-clamping modes \cite{shin_spiderweb_2022} demonstrate how tailored mechanical designs can dramatically reduce dissipation, thereby improving stability and signal control. Despite this progress, the reliance on conventional nanofabrication methods can limit the iterative design process needed to optimize mechanical performance and explore new SiN resonator geometries.

Conventional fabrication of SiN resonators involving lithography and multiple etching steps can be inflexible and time-consuming, while inevitably resulting in structures with overhanging material that increases mechanical dissipation. Photolithography relies on photomasks with fixed patterns, making iterative prototyping tedious, particularly when in-house photomask fabrication is not possible. While e-beam lithography offers a more flexible, maskless alternative, writing millimeter-scale features can take tens of hours \cite{greve_optimization_2013,li_high_2016}. In many cases, two front-to-back aligned etch masks must be lithographically patterned—one defining the SiN resonant structure and the other a backside window for subsequent structural release—introducing additional complexity, processing time, and misalignment error \cite{reinhardt_ultralow-noise_2016,norte_mechanical_2016}. Typically, these lithographic patterns are transferred to the SiN film by reactive ion etching (RIE), after which an anisotropic wet (e.g., KOH) or dry (e.g., RIE with SF$_6$) etch of the silicon substrate is used to release the SiN structure. Despite the high selectivity of KOH to silicon crystal planes, the extended etch times required for bulk material removal (tens of hours \cite{reinhardt_ultralow-noise_2016}) leads to some lateral etching of slower-etching planes, causing unintended undercut of the supports. This undercut also occurs with dry etching \cite{shin_spiderweb_2022} and, in both cases, leads to overhang that degrades the quality factor of the fabricated resonator \cite{reinhardt_ultralow-noise_2016,norte_mechanical_2016,bereyhi_hierarchical_2022}. Moreover, errors in front-to-back lithography alignment can exacerbate overhang at certain locations on the final structure, thereby worsening mechanical dissipation.

Previous work has demonstrated that nanomechanical resonators, including nanobeams and trampolines, can be rapidly fabricated by direct femtosecond laser ablation of free-standing SiN membranes \cite{xie_laser_2023,nikbakht_high_2023}. Others have used ultrafast laser ablation of SiN membranes to create micron- and nanometer-scale through-holes \cite{bonse_femtosecond_2013,uesugi_nanoprocessing_2021}, nanopores \cite{leva_localized_2025}, slit arrays \cite{uesugi_ultrafast_2020}, and multi-hole lattices \cite{uesugi_multi-beam_2023}---the latter two achieved through the interference of multiple Gaussian beams. However, fabricating non-periodic complex geometries has posed significant challenges, as prestressed SiN membranes are prone to cracking during laser machining. In \cite{xie_laser_2023}, continuously firing the laser along a path was found to induce cracks in SiN membranes due to stress concentrations that developed at the path tip once a critical length was reached. Instead, microholes were sequentially ablated along the perimeter of each shape, eventually overlapping to release the material while limiting local stress, thereby preventing membrane failure \cite{xie_laser_2023,nikbakht_high_2023}.

While the techniques developed in \cite{xie_laser_2023,nikbakht_high_2023} showed great potential, the complexity of structures that could be fabricated was constrained by the need to manually program a translation stage and laser before machining each new design. This involved writing numerical control code that combined the ablation routines for simple shapes to construct the desired cutouts over many intermediate steps. For instance, nanobeams were fabricated by first machining an array of circles which were later connected with straight lines \cite{nikbakht_high_2023}---a method that cannot be easily extended to create the more intricate geometries found in many state-of-the-art resonators \cite{shin_spiderweb_2022,bereyhi_hierarchical_2022}. Consequently, although laser machining itself could be completed in minutes, the manual preparation process was much more time-consuming, rendering the approach impractical for rapid prototyping of diverse device architectures.

Here we present a software toolset that generates laser-ablated hole sequences directly from standard semiconductor layout files (i.e., GDSII format), enabling crack-free fabrication of SiN resonators with arbitrary geometries. Included in this software is an alignment tool that limits residual SiN overhang to $\sim2$ \textmu m, minimizing the associated Q-factor degradation \cite{reinhardt_ultralow-noise_2016,norte_mechanical_2016,bereyhi_hierarchical_2022}. We use the toolset to fabricate a variety of trampoline-based resonators, including a branched-clamp geometry that achieves a Q-factor of $3.7 \times 10^6$ at 60 kHz. Comparison with finite element method (FEM) simulations indicates that these structures have material Q-factors above 3700---i.e., within the range for plain SiN membranes \cite{nikbakht_high_2023,villanueva_evidence_2014,schmid_fundamentals_2023}---suggesting that our laser machining process does not significantly affect intrinsic material dissipation.
	
\section{Methods}
Structures are fabricated by mounting a SiN membrane chip on a translation stage that moves beneath a pulsed laser beam, with the stage and laser programmed to ablate a sequence of holes that cut out the desired geometry (see Fig. 1). Free-standing square silicon-rich SiN membranes, measuring 1.2 or 1.7 mm in side length and 100 nm in thickness, serve as substrates for the laser machining process. The membranes are produced in-house using contact photolithography and KOH etching \cite{mu_remote_2023} but could also be purchased commercially from various suppliers. The fundamental resonant frequency $f_{1,1}$ of a sample membrane is measured and used to extract a residual stress value of $\sim76$ MPa using the analytical model in \cite{schmid_fundamentals_2023}: $f_{1,1}=1/\sqrt{2}L\sqrt{\sigma/\rho}$, where $L$ is the square membrane side length, $\sigma$ is the residual stress, and $\rho$ is the density of SiN (2900 kg/m$^3$). Chips are mounted on an Aerotech nano-precision 3-axis stage using double-sided tape, with the z-axis fixed prior to machining to achieve optical focus. Laser machining is performed using a femtosecond laser system (PHAROS from Light Conversion) operating at a fundamental wavelength of 1030 nm, halved to 515 nm using a barium borate crystal. We use a pulse duration of 320 fs and a repetition rate of 200 kHz. The laser beam is focused using a 20 $\times$ infinity corrected high working distance NIR Mitutoyo microscope objective with a numerical aperture of 0.4, achieving a theoretical spot size of 1 \textmu m. A power meter is used to measure and adjust the average laser power to 10 mW, corresponding to a pulse energy of 50 nJ. Each hole is produced using 3 pulses. The entire system is controlled by an Aerotech A3200 machine controller, programmed with a unique numerical control file for each structure to be machined. Additional details and schematics of the femtosecond laser machining setup are available in \cite{xie_laser_2023,nikbakht_high_2023}.

\begin{figure*}[htbp]
	\centering
	\includegraphics[width=\textwidth]{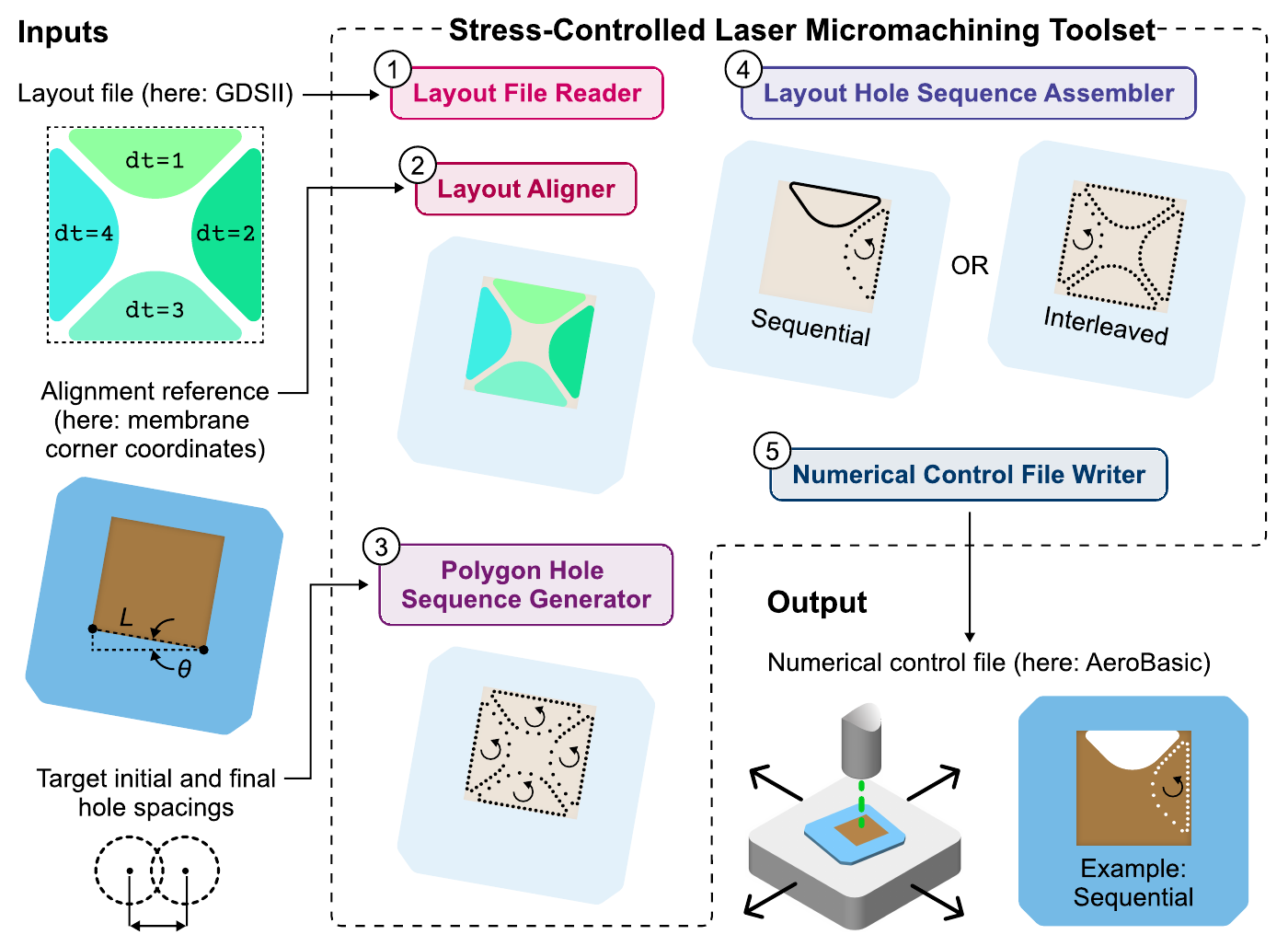}
	\caption{\textbf{Schematic of the Stress-Controlled Laser Machining Toolset (SCLMT): a software library comprising five tools connected through a pipeline.} (1) Layout File Reader extracts geometric elements from a layout file and converts them into polygons, (2) Layout Aligner transforms the layout to align with the substrate, (3) Polygon Hole Sequence Generator creates a coordinate list defining the sequence of laser-ablated holes used to cut out an individual polygon (see also Fig. 2), (4) Layout Hole Sequence Assembler combines all polygon hole sequences into a single layout-wide sequence, and (5) Numerical Control File Writer translates the layout hole sequence into a numerical control file for a translation stage and laser. Depictions of laser-ablated holes are not to scale.}
\end{figure*}

We develop an open-source Python package that automatically converts conventional semiconductor layout files into numerical control code for laser micromachining (schematized in Fig. 1). This package, referred to as the Stress-Controlled Laser Micromachining Toolset (SCLMT), comprises a modular pipeline of five tools, each performing a distinct processing step. At the core of the system is the Polygon Hole Sequence Generator (PHSG), which determines where and in what order laser-ablated holes must be placed to cut out an arbitrary shape without cracking the membrane. Importantly, the modular architecture of the pipeline allows users to substitute any tool (except the PHSG) with custom implementations---such as a Numerical Control File Writer for a specific motion and laser controller---without modifying the rest of the toolset. This flexibility is achieved by defining each tool as an abstract class and using dependency injection throughout the pipeline. The source code and documentation are publicly available at \cite{mems-laser-machining}. The remainder of this section describes the five specific tool implementations used in this work.

We implement a Layout File Reader (Tool 1) that parses GDSII files containing the resonator layouts to be fabricated. Here, GDSII files are prepared similarly to conventional semiconductor layouts, allowing existing photomask designs to be easily repurposed for laser machining. The only deviation from standard photomask drawing practices is the use of the \texttt{datatype (dt)} attribute to encode the machining order of each geometric element (see layout input in Fig. 1). All geometries are otherwise assigned to a single layer within a flat cell. The tool approximates all geometry, including rounded shapes and paths, as polygons, representing each as an array of vertices.

Our implementation of the Layout Aligner (Tool 2) scales and rotates the layout to precisely fit with a stage-mounted membrane (see Fig. 1.2), allowing structures to be fabricated with at most 2 \textmu m of residual SiN overhang. The coordinates of two membrane corners are measured and used to calculate the angular misalignment between the stage and membrane as well as the actual membrane size, which is often slightly different than assumed in the layout file. Although unused in this work, alternative implementations of the Layout Aligner could employ any other alignment marks instead of the membrane corners and apply arbitrary affine transformations to the layout.

Here, alignment measurements are performed by manually jogging the stage to align a crosshair (superimposed on a camera image from a microscope along the laser beam path) with the bottom left and right corners of the membrane. Day-to-day disturbances to the optics introduce a small offset between the crosshair and the true laser focus spot, which we compensate before each run by ablating a single test hole and determining the offset between the crosshair and hole center. Crosshair-based measurements of corner and hole positions are limited by the microscope magnification, camera resolution, and crosshair width, resulting in an uncertainty roughly equal to a quarter of the hole diameter: $\pm250$ nm in our case. Combining the three crosshair-based measurement uncertainties (1 for the offset, and 2 for the corners) with the stage accuracy (200 nm) and repeatability (75 nm) yields a total uncertainty of roughly $\pm1$ \textmu m in hole positions relative to the membrane border. To ensure that all holes remain fully within the membrane under worst-case conditions, we apply a 1 \textmu m padding to the measurements provided to the Layout Aligner, resulting in a maximum possible overhang of $\sim2$ \textmu m.

\begin{figure}[htbp]
	\centering
	\includegraphics[width=\columnwidth]{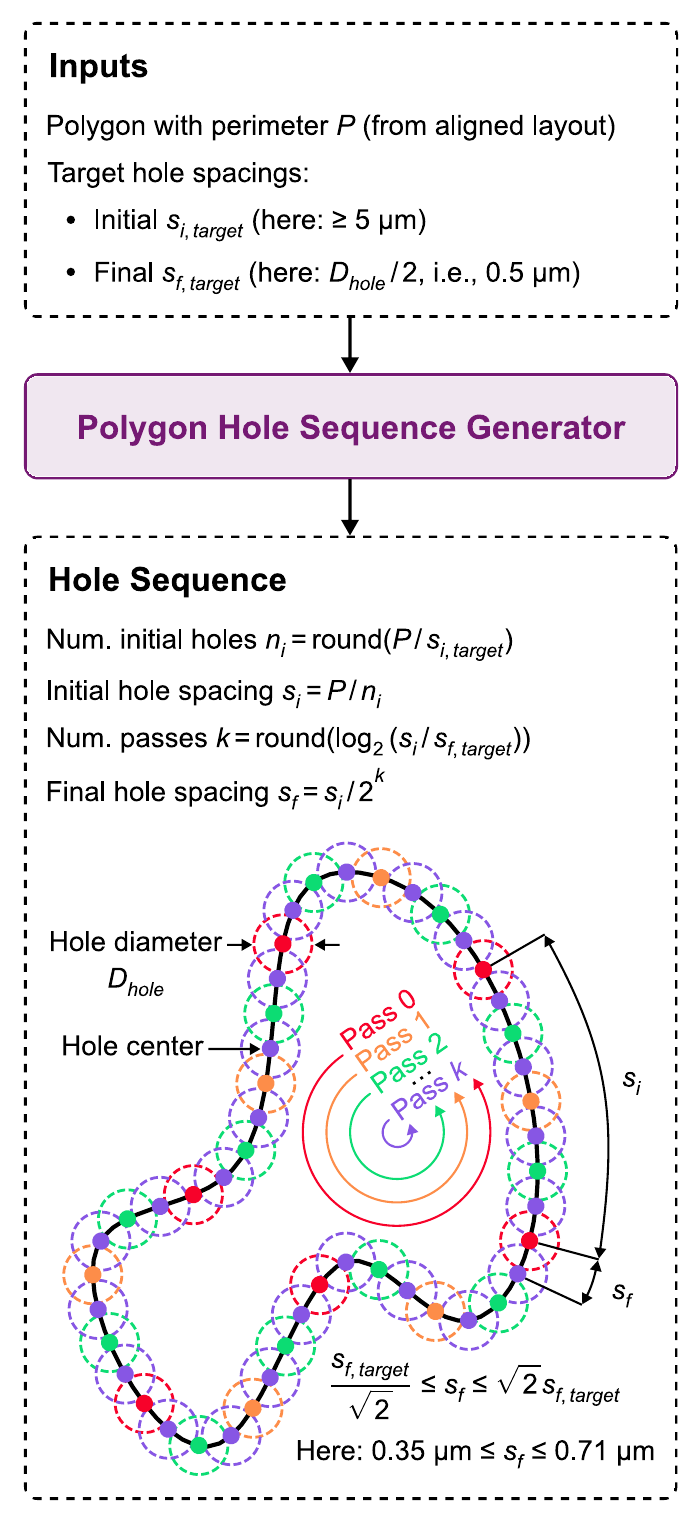}
	\caption{\textbf{Schematic of the Polygon Hole Sequence Generator (Tool 3 of SCLMT)}. This tool takes a polygon from the aligned layout as input, along with user-defined targets for the initial and final pass spacings between adjacent holes. From these inputs, the number of initial holes, actual initial spacing, number of passes, and actual final spacing are computed. Color coding shows how holes are placed in successive passes along the polygon perimeter, progressively halving the spacing until holes overlap and the polygon is released. The initial pass (pass 0) is shown in red; the final pass (pass $k$) in purple.}
\end{figure}

The Polygon Hole Sequence Generator (Tool 3; Fig. 1.3 and Fig. 2) outputs a list of coordinates defining the sequence in which holes are laser-ablated to cut out a polygon while preventing membrane fracture. To limit local stress, holes are placed in successive passes around the perimeter, with each pass halving the spacing until holes eventually overlap and release the shape. Polygons from the aligned layout are processed independently, each through a separate call to the PHSG. Users configure this tool with the target hole spacings at the end of the initial and final passes, denoted $s_{i,target}$ and $s_{f,target}$, respectively. Here, spacing is defined as the distance between the centers of adjacent holes along the polygon perimeter.

In the initial pass, we place $n_i=\text{round}(P/s_{i,target})$ equally spaced holes along the polygon perimeter $P$, giving an initial spacing of $s_i=P/n_i$. In each subsequent pass, a hole is placed at the midpoint of every segment between adjacent holes, halving the spacing. After $k$ passes excluding the initial one, the final spacing becomes $s_f=s_i/2^k$---alternatively expressed as $s_f=P/N$, where $N=n_i2^k$ is the total number of holes. To approximate the target final spacing, we set $k=\text{round}(\text{log}_2(s_i /s_{f,target}))$, which bounds $s_f$ in the interval $\left[\frac{1}{\sqrt{2}}s_{f,target}, \sqrt{2}s_{f,target}\right]$.

We ensure adjacent holes overlap in the final pass by setting the target final spacing to half the hole diameter---i.e., $s_{f,target}=$ 0.5 \textmu m for the 1 \textmu m-diameter holes used here \cite{xie_laser_2023}---resulting in $s_f$ values between 0.35 and 0.71 \textmu m, all of which consistently yield structural release. Smaller values of $s_f$ tend to reduce edge roughness, albeit at the cost of longer machining times.

A complementary constraint applies to the initial spacing, which must be small enough—relative to the polygon perimeter—to allow for at least two initial holes (i.e., $s_i\leq P/2$), but also large enough to avoid early membrane failure; in practice, we find 5 \textmu m to be a suitable lower bound. Users may manually set $s_{i,target}$ to any value within this valid range or allow the SCLMT to automatically select a value that minimizes the deviation between $s_{f,target}$ and $s_f$.

We implement two versions of the Layout Hole Sequence Assembler (LHSA, Tool 4)---sequential and interleaved—that both assemble individual polygon hole sequences into a layout-wide sequence but offer different ways of controlling the evolution of built-in stress during machining. Sequential execution fully releases a polygon before machining the next, whereas interleaved execution completes pass $i$ of each polygon before starting pass $i+1$ of the first (see Fig 1.4). Certain structures with large stress concentrations, such as those shown in Fig. 3(c,e) are observed to crack when sequential execution is used. This is attributed to the removal of large sections of the membrane early in the process which causes tensile or biaxial forces to redistribute over a much smaller cross-sectional area. Interleaved execution is designed to more equally distribute stress during machining and allows for the repeatable fabrication of such structures; however, it requires greater stage displacement, making it more time-consuming than sequential execution.

Finally, our Numerical Control File Writer (Tool 5) converts layout hole sequences into AeroBasic files needed to programmatically translate the stage and fire laser pulses (see output in Fig. 1). These files use G-code commands to control stage motion and position synchronized output to trigger laser pulses based on the stage position \cite{aerotech_position}.

Details on cleaning procedures, as well as experimental characterization and FEM simulation methods used to evaluate the mechanical response of fabricated structures are provided in Supplementary Sections S1, S2 and S4, respectively.

\section{Results}
\begin{figure}[htbp]
	\centering
	\includegraphics[width=\columnwidth]{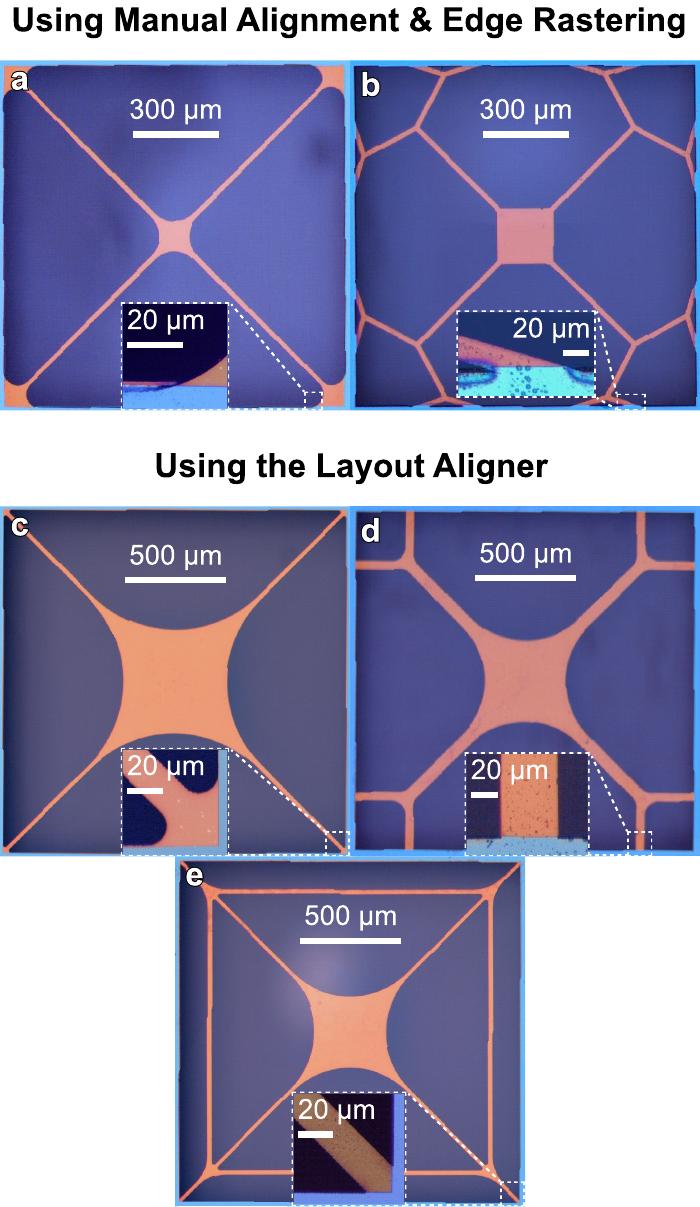}
	\caption{\textbf{Trampoline-based resonant structures fabricated using the SCLMT.} Each device is machined from a free-standing square SiN membrane with a side length of either 1.2 mm or 1.7 mm. Structures (c–e) are fabricated using the Layout Aligner (Tool 4 of SCLMT), whereas (a,b) are fabricated using an earlier method involving coarse manual alignment and a corrective raster scan around the border-lying edges of each polygon to ensure structural release. (a) 1.2 mm trampoline, (b) 1.2 mm hierarchical trampoline, (c) 1.7 mm trampoline, (d) 1.7 mm branched-clamp trampoline, and (e) 1.7 mm webbed trampoline.}
\end{figure}

Figure 3 presents several trampoline-based resonant structures with varying geometries and feature sizes, fabricated using the SCLMT. The smallest features we fabricate are the 15 \textmu m-wide tethers of the trampoline in Fig. 3a; however, this is not necessarily the minimum feature size achievable with laser machining, since beams as narrow as 7 \textmu m have been demonstrated in previous work \cite{nikbakht_high_2023}. The structures in Fig. 3(c,e) require the use of the interleaved LHSA (Tool 4 of SCLMT; see Fig. 1.4) and would otherwise crack during machining due to their high-aspect-ratio features. Machining times range from 20 minutes to 1 hour, with the larger and more complex geometries generally taking longer to fabricate.

Importantly, structures fabricated using the Layout Aligner (Tool 2 of SCLMT; see Fig. 1.2) exhibit less than 2 \textmu m of overhang, as shown in the insets of Fig. 3(c-e). As described in the Methods section, the overhang is limited by the magnification and resolution of the imaging system used for alignment measurements, as well as the accuracy and repeatability of the stage displacement. In contrast, the structures in Fig. 3(a,b) are fabricated using an earlier method, where membranes are coarsely aligned by hand, and a final corrective step---raster-scanning the laser around the border-lying edges of each polygon---is performed to ensure structural release. We abandoned this brute-force technique as it ablated the underlying silicon, generating large amounts of debris that redeposited onto the final structures. Only structures fabricated using the Layout Aligner are experimentally characterized in this section.

Most of these structures are fabricated without exhaustive prior FEM simulations, with the primary goal of benchmarking the fabrication process. A secondary aim is to identify promising resonator geometries for radiation absorption and sensing---specifically, structures that are coupled to their environment primarily through thermal radiation and exhibit high quality factors \cite{snell_heat_2022,zhang_high_2024}. To this end, trampolines are chosen as a base structure for their high thermal isolation \cite{piller_thermal_2023} and are augmented with soft-clamping geometries that others have demonstrated can decrease mechanical dissipation \cite{bereyhi_hierarchical_2022,shin_spiderweb_2022}.

The conventional trampoline in Fig. 3c is mainly designed to maximize thermal isolation through the inclusion of a large central structure, relatively thin tethers (20 \textmu m-wide), and minimal clamp-widening. This structure exhibits a maximum Q-factor of $2.2\times10^6$ at 27 kHz. FEM simulations indicate that 89\% of its total heat transfer with the surrounding environment occurs through radiation---the highest radiative fraction among all fabricated designs. This estimate is based on the simulation method outlined in \cite{zhang_enhanced_2025}, in which a heat load of $Q_{tot}=$ 1 \textmu W is uniformly applied to the resonator surface. The resulting total conductive heat flux through all four tethers is found to be $Q_{cond}=$ 110 nW, yielding a radiative contribution of $x_{rad}=(Q_{tot}-Q_{cond})/Q_{tot}\approx89\%$.

The hierarchical structure shown in Fig. 3b, inspired by \cite{bereyhi_hierarchical_2022}, is intended to minimize bending at the silicon clamping points compared to a conventional trampoline. We do not experimentally characterize this design since subsequent FEM simulations reveal that it would actually yield a low Q-factor relative to a traditional trampoline. This poor performance arises from a lack of self-similarity in the branched geometry---unlike in \cite{bereyhi_hierarchical_2022}---which leads to unequal stress at the clamps, thereby degrading dissipation dilution. To address this issue, we design the branched-clamp trampoline shown in Fig. 3d, which achieves a maximum Q-factor of $3.7\times10^6$ at 60 kHz---the highest among the fabricated structures. Later FEM simulations indicate that this branching scheme yields a Q-factor comparable to that of conventional trampolines with similar dimensions (see Supplementary Section S5). Nevertheless, the branched-clamp design offers a marginal advantage by enabling the use of the longest possible tethers (i.e., diagonally oriented) while still being anchored perpendicularly to the silicon edge. Anchoring perpendicularly, rather than at the four corners as in conventional trampolines, opens the possibility of investigating clamp-tapering to reduce clamping loss, which was only proven successful with perpendicular clamping \cite{bereyhi_clamp-tapering_2019}.  Additional design details and performance characterization of the branched-clamp trampoline can be found in \cite{zhang_enhanced_2025}, where this particular resonator is shown to operate at the fundamental performance limit of radiation sensing.

Drawing on the spiderweb resonator from \cite{shin_spiderweb_2022}, we design the structure shown in Fig. 3e, which is predicted to have two high-quality-factor soft-clamping modes localized in the beams connecting the diagonal tethers. However, we were unable to excite these modes during characterization; instead, we observe a maximum Q-factor of $2.7\times10^6$ at the fundamental eigenfrequency of 15 kHz. Unlike the original structure in \cite{shin_spiderweb_2022}, the soft-clamping modes are not designed to be well-isolated from the eigenfrequencies of nearby piston modes, which likely led to hybridization that suppressed their response during characterization. This structure remains a work in progress; we continue to iterate on the design, leveraging the flexibility of the SCLMT to rapidly prototype new geometries.

\begin{figure}[htbp]
	\centering
	\includegraphics[width=\columnwidth]{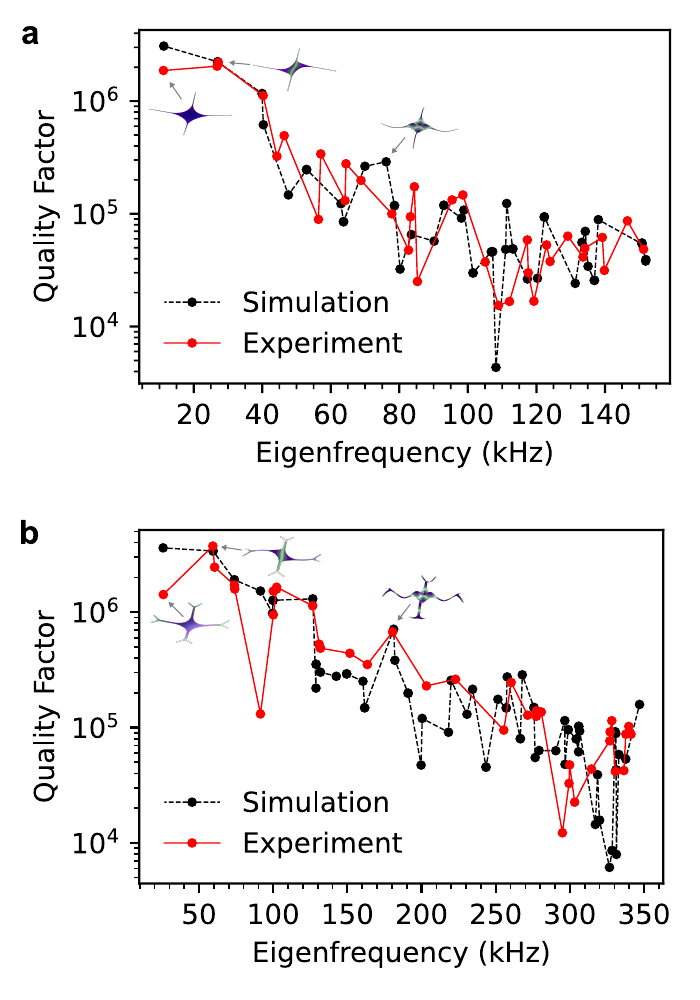}
	\caption{\textbf{Experimental characterization of the resonance frequencies and quality factors} of (a) the conventional trampoline shown in Fig. 3(c,b) the branched-clamp trampoline shown in Fig. 3d. Material quality factors ($Q_{mat}$) of 4000 and 3700 are extracted for the conventional and branched-clamp trampolines, respectively, by fitting finite element simulations of dissipation dilution to the experimental quality factor measurements. The simulation points, shown in red, represent the product of the simulated dissipation factors and the extracted $Q_{mat}$ values.}
\end{figure}

We extensively characterize the mechanical response of the conventional and branched-clamp trampolines in Fig. 3(c,d) to assess whether laser machining introduces additional material dissipation (i.e., reduces the material Q-factor $Q_{mat}$). Figure 4 presents the measured Q-factors of many modes compared to those predicted by FEM simulations for both structures. We assume the Q-factor is dominated by material dissipation and can thus be approximated as $Q\approx\alpha_{dd}Q_{mat}$, where $\alpha_{dd}$ is the dissipation dilution factor. For our chamber pressure of $1.2\times10^{-6}$ hPa, we estimate vacuum-limited Q-factors of $\sim15\times10^6$ and $\sim36\times10^6$ at the fundamental eigenfrequencies of 11 and 26 kHz for the conventional and branched-clamp trampolines, respectively---ruling out gas damping as a significant loss mechanism (see Supplementary S3). Fitting the simulated dissipation dilution to the measured Q-factors yields $Q_{mat}$ values of 4000 and 3700 for the conventional and branched-clamp trampolines, respectively. Additionally, we observe close agreement between the simulated and measured frequency dependence of the Q-factor for both structures.

The obtained $Q_{mat}$ estimates fall within the expected range for a 100 nm-thick SiN plain membrane as measured in our previous work (e.g., $Q_{mat}\approx$ 2700 \cite{nikbakht_high_2023}) and in the literature ($Q_{mat}\approx$ 5700 in \cite{villanueva_evidence_2014}), suggesting that our laser machining process has no significant effect on material dissipation. This is in contrast to previous work, wherein laser-ablated regions of SiN were shown to be silicon-rich \cite{xie_laser_2023}, which noticeably degraded $Q_{mat}$ \cite{nikbakht_high_2023}. However, unlike this previous work, we do not perform a final cleaning pass in which the laser is continuously pulsed along the perimeter of each shape \cite{xie_laser_2023,nikbakht_high_2023}. This omission greatly reduces the total laser energy incident on the SiN, likely minimizing the impact on the chemical composition of the machined material. Lastly, our use of structures with clamps wider than 20 \textmu m likely reduces the fraction of affected material at the clamps, helping curtail any negative effects on dissipation dilution.

\section{Conclusion}
We demonstrate laser machining as a powerful platform for rapid prototyping of SiN nanomechanical resonators---offering a level of flexibility and speed unattainable with conventional nanofabrication. The development of a Layout Aligner enables fabrication of structures with at most 2 \textmu m of residual SiN overhang, which could likely be further reduced using alignment imaging equipment with higher magnification and resolution. With the introduction of an interleaved Layout Hole Sequence Assembler, we fabricate high-aspect-ratio structures with large stress concentrations, broadening the design space accessible through laser machining. Finally, we show that our process does not noticeably degrade material dissipation, yielding devices with performance comparable to those produced through conventional techniques.

In future work, we aim to extend this platform to high-stress SiN membranes (e.g., $\sim1$ GPa), which are widely used for their superior material quality and dissipation dilution. Successfully adapting our laser machining technology to these membranes would enable the fabrication of even higher-performance devices and further establish femtosecond ablation as a robust and versatile method for nanomechanical resonator prototyping.

\bibliographystyle{apsrev4-2-titles}
\bibliography{references}

\clearpage

\section*{S1. Cleaning procedure}
Occasionally, cutouts generated during laser machining are redeposited elsewhere on the membrane upon release, particularly in layouts with small features. Luckily, structures where most of the membrane area has been machined, such as trampolines, can be manually agitated to remove scraps without risking fracture. No further cleaning procedure has been investigated in the context of this work. 

\section*{S2. Mechanical response characterization}
Mechanical quality factors of the fabricated structures are characterized in a custom high-vacuum chamber ($\sim7.5\times10^{-7}$ Torr). The Q-factors are measured using a fiber-optic interferometric setup designed to measure out-of-plane vibrations \cite{snell_heat_2022}. The setup consists of a cleaved single-mode optical fiber positioned in front of the resonator at normal incidence, forming a Fabry–Pérot cavity between the fiber tip and the structure surface. Light from a 1550 nm Orion$^{\text{TM}}$ laser is coupled using a 90:10 single mode (mode-field diameter of $\sim10$ \textmu m) into the fiber and partially reflected both at the cleaved fiber end and at the structure surface \cite{rugar_improved_1989}. Interference between these two reflections produces a signal that is sensitive to nanometer-scale displacements of the resonator. The interferometric signal is detected using a photodetector (Thorlabs Inc. PDA20CS2) and recorded over time using a Zurich Instrument Ltd. MFLI lock-in amplifier to observe the free decay of the resonator’s motion. The Q-factor is extracted using the ringdown technique, where the decay envelope of the oscillation amplitude is fitted to an exponential function.

\section*{S3. Vacuum-limited Q-factors}
We estimate the vacuum-limited Q-factors of the conventional and branched-clamp trampolines in Fig. 3(c,b) using the analytical model in \cite{schmid_fundamentals_2023} for drag-force damping of an oscillating plate in the ballistic regime: 

\renewcommand{\theequation}{S\arabic{equation}}
\begin{equation}
	Q_{b\text{--}df} = \frac{\rho h \omega}{4} 
	\sqrt{\frac{\pi}{2}} 
	\sqrt{\frac{R_{\text{gas}} T}{M_m}} 
	\frac{1}{p},
\end{equation}

\noindent where $\rho$ is the density of SiN (2900 kg/m$^3$), $h$ is the thickness (100 nm), $\omega$ is the angular velocity of oscillation for the fundamental mode ($2\pi\times11\times10^3$  rad/s for Fig. 3c, $2\pi\times26\times10^3$  rad/s for Fig. 3d), $R_{\text{gas}}$ is the universal molar gas constant (8.314 J/mol$\cdot$K), $T$ is the temperature (293.15 K), $M_m$ is the molar mass of dry air (28.97 g/mol), and $p$ is the pressure ($1.2\times10^{-6}$  hPa).

\vspace{2em}
\section*{S4. Dissipation dilution FEM simulation}
To assess the predictability of mechanical performance in laser-machined structures, we perform finite element method (FEM) simulations using COMSOL Multiphysics. Here, each structure is modelled as a SiN shell with a thickness $h=$ 100 nm, Young’s modulus $E=$ 285 GPa, Poisson’s ratio $\nu=$ 0.25, and a density $\rho=$ 2900 kg/m$^3$. We apply an isotropic external stress of 76 MPa, accounting for the membrane stress prior to machining, and impose fixed boundary conditions at the edges of the structure which are anchored to the silicon substrate. We use an unstructured quadrilateral mesh with a higher element density along the fixed edges to accurately capture clamping curvature and thus bending losses at the clamping points. Through a mesh convergence analysis, we find that maximum and minimum element sizes of 5 \textmu m and 0.34 \textmu m, respectively, and a distribution of $\sim1.4$ elements per micron along the fixed edges, yield accurate results.

Mechanical performance is simulated through a two-step prestressed eigenfrequency study which first computes the relaxed stress state of the structure, followed by the frequency and shape of each vibrational mode. To facilitate a comparison with experimental Q-factor measurements, we find the dissipation dilution factor $\alpha_{dd}$ using the ratio of kinetic and linear elastic energies by computing the following surface integral–adapted from \cite{bereyhi_hierarchical_2022}---over each mode shape:

\begin{widetext}
\begin{equation}
	\alpha_{dd} = \frac{12\rho \omega_n^2}{E h^2}(1 - \nu^2)
	\frac{
		\displaystyle \int w^2 \, dx\,dy
	}{
		\displaystyle \int \left[
		\left( \frac{\partial^2 w}{\partial x^2} + \frac{\partial^2 w}{\partial y^2} \right)^2
		+ 2(1 - \nu) \left(
		\left( \frac{\partial^2 w}{\partial x \partial y} \right)^2
		- \frac{\partial^2 w}{\partial x^2} \frac{\partial^2 w}{\partial y^2}
		\right)
		\right] dx\,dy
	},
\end{equation}
\end{widetext}

\noindent where $\omega_n$ is the eigenfrequency, and $w$ is the out-of-plane component of the displacement field.

\section*{S5. Branched-clamped trampoline benchmarking simulation}
\label{S5}

\renewcommand{\thefigure}{S\arabic{figure}}
\setcounter{figure}{0}
\begin{figure*}[htbp]
	\centering
	\includegraphics{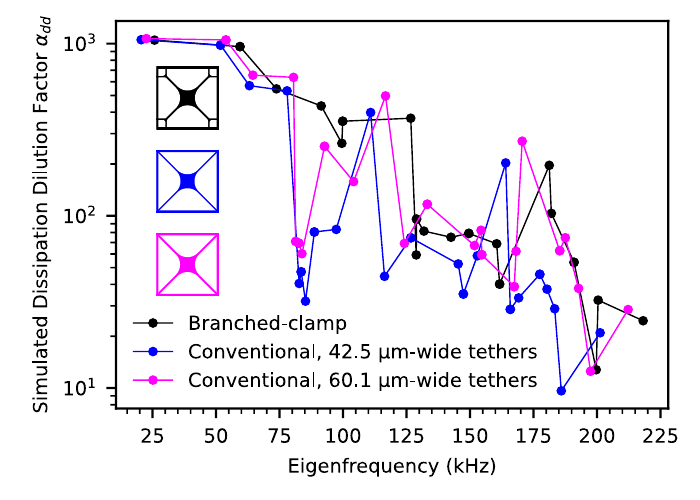}
	\caption{Simulated dissipation dilution of the branched-clamp trampoline in Fig. 3d (black), a conventional trampoline with 42.5 \textmu m-wide tethers (blue), and a conventional trampoline with 60.1 \textmu m-wide tethers (magenta).}
\end{figure*}

The branched-clamp trampoline shown in Fig. 3d is benchmarked against two conventional trampolines. One has tethers matching the width of the source tether in the branched-clamp design (42.5 \textmu m), while the other uses tethers equal in width to the branched tethers (60.1 \textmu m). The conventional trampolines do not feature any clamp-widening and anchor directly into the corners. All three designs share an identical central structure. As shown in Fig. S1, FEM simulations indicate that the dissipation dilution---and therefore the Q-factor---is similar across all three structures over a broad range of eigenfrequencies.

\end{document}